\documentclass{aa}
\usepackage{natbib}
\usepackage{graphicx}
\usepackage[varg]{txfonts}

\begin{document}

\title{Nonthermal emission properties of the northwestern rim of supernova remnant RX~J0852-4622}

\author{Tetsuichi Kishishita\inst{1}
  \and Junko Hiraga\inst{2} 
 \and Yasunobu Uchiyama\inst{3, 4}} 


\institute{Universiy of Bonn, Physikalisches Institut, Nussallee 12, Bonn 53115, Germany
  \and  Department of Physics, The University of Tokyo, Bunkyo, Tokyo 113-0033, Japan
  \and Kavli Institute for Particle Astrophysics and Cosmology, SLAC National Accelerator 
Laboratory 2575 Sand Hill Road M/S 29, Menlo Park, CA 94025
\and Panofsky Fellow
}

\date{Received 09 October 2012 / Accepted 25 January 2013}

\abstract{
The supernova remnant (SNR) RX J0852-4622 (Vela Jr., G266.6-1.2) is one of the most important SNRs for investigating the acceleration of multi-TeV particles and the origin of Galactic 
cosmic rays because of its strong synchrotron X-ray and TeV $\gamma$-ray emission, which show a shell-like 
morphology similar to each other. Using the {\it XMM}-{\it Newton} archival data consisting of multiple 
pointing observations of the northwestern rim of the remnant, we investigate the spatial properties of 
the nonthermal X-ray emission as a function of distance from an outer shock wave. All X-ray 
spectra are well reproduced by an absorbed power-law model above 2 keV. It is found that the 
spectra show gradual softening from a photon index $\Gamma=2.56$ in the rim region to $\Gamma=2.96$ in the interior region. We show that this radial profile can be interpreted as a gradual decrease of the cutoff energy of the electron spectrum due to synchrotron cooling. By using a simple spectral evolution model that includes continuous synchrotron losses, the spectral softening can be reproduced with the magnetic field strength in the post-shock flow to less than several tens of $\mu$G. If this is a typical magnetic field in the SNR shell, $\gamma$-ray emission would be accounted for by inverse Compton scattering of high-energy electrons that also produce the synchrotron X-ray emission. Future hard X-ray imaging observations with {\it Nustar} and ASTRO-H and TeV $\gamma$-ray observations with the Cherenkov Telescope Array (CTA) will allow to us to explore other possible explanations of the systematic softening of the X-ray spectra.} 

\maketitle 

\section{Introduction}
Supernova remnants (SNRs) have been considered as major cosmic-ray (CR) accelerators below 
the {\it knee} energy of $\sim10^{15}$~eV. The measured energy density of CRs can be explained if $\sim10\%$ of each supernova kinetic energy is transferred to the accelerated particles. A plausible mechanism for this particle acceleration is diffusive shock acceleration (DSA; e.g., \citealt{bell, drury, blandford}).
In the DSA theory, particles are accelerated at the shock, resulting 
in a power-law distribution, which can account for the Galactic CR spectrum using a reasonable 
diffusion coefficient in the Galaxy. However, many unsolved problems are still remaining, such as 
the acceleration efficiency, the maximum energy of the accelerated particles, and so on. Observations 
of synchrotron X-ray emission have been playing important roles in stimulating the development 
of the DSA theory. For example, magnetic field amplification at the shocks of young SNRs has 
been inferred from X-ray observations (e.g., \citealt{uchiyama}) and it is now considered as an 
integral part of the DSA theory (e.g., \citealt{ellison}).

RX~J0852.0-4622 (also called G226.2-1.2 or Vela Jr.) is a young SNR with $\sim2^{\circ}$ of an angular size in the line of sight to the Vela SNR. RX~J0852.0-4622 was originally discovered in the data of the {\it ROSAT} All-Sky Survey (\citealt{asch}). Based on the {\it ROSAT} data, whose bandpass was limited to soft X-rays below 2 keV, \cite{asch} argued that the emission from the SNR can be explained either by a hot thermal model with a temperature of 
$\sim$2.5~keV or by a power-law model with a photon index of $\Gamma\sim2.6$. {\it ASCA} observations, with an imaging capability of up to $\sim$10~keV, showed that the X-ray emission from the bright rims of RX~J0852.0-4622 is dominated by a nonthermal emission characterized by a power-law with a photon index $\Gamma\sim2.6$ (\citealt{tsunemi, slane}). The nonthermal X-ray emission is presumably of synchrotron origin. 
The {\it Chandra} image of the northwestern (NW) rim revealed sharp filamentary structures similar to 
those discovered in SN1006 (\citealt{bamba}).

The CANGAROO collaboration claimed the detection of TeV $\gamma$-rays from the direction that 
coincides with the peak of the X-ray emission in the NW rim \citep{katagiri}. Then, the 
H.E.S.S. telescopes, with their highly sensitive stereoscopic observations, have detected spatially 
extended TeV $\gamma$-ray emission, which shows a good spatial correlation with the nonthermal X-ray 
shells \citep{aha07}. The H.E.S.S. spectrum of RX~J0852.0-4622 can be well fitted with 
a power-law of $\Gamma\sim2.1$. Recently, the Large Area Telescope (LAT) on board the {\it Fermi} Gamma- 
ray Space Telescope has detected $\gamma$-ray emission from RX~J0852.0-4622 in an energy band of 
1–300 GeV \citep{tanaka11}. The LAT spectrum is described by a power-law with a photon index $\Gamma\simeq1.85 \pm 0.06$, which is substantially harder than the H.E.S.S. spectrum. The origin 
of the $\gamma$-ray emission has been actively debated; the GeV–TeV $\gamma$-rays can be explained either by 
inverse Compton (IC) scattering of high-energy electrons, or by $\pi_0$-decay $\gamma$-rays produced in interactions of accelerated protons (and nuclei) with interstellar gas \citep{aha07}. SNR RX~J0852.0-4622 is one of the most interesting objects for studying CR acceleration at SNR shocks.

\begin{figure*}[htdp]
      \includegraphics[width=15cm]{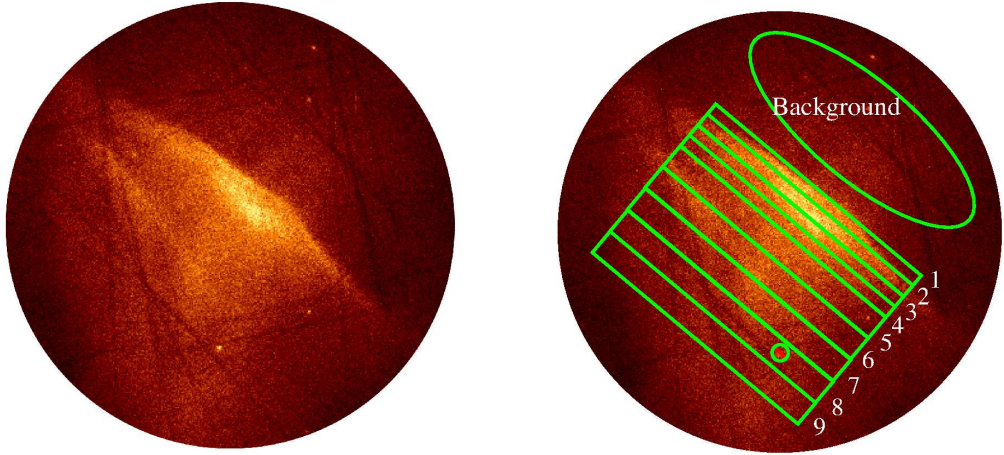}
  \caption{EPIC-MOS image of the northwestern rim of RX~J0852.0 -4622 in an energy range of 0.5–10 keV 
(left) and the scheme to search for the radial profile of the synchrotron X-ray emission (right). The 
angular size of the summed image is larger than {\it XMM}-{\it Newton}’s FOV. Numbers in figure indicate region 
IDs. A contamination source in region 8 was removed from the spectral analysis. }
   \label{fig1} 
\end{figure*}

The age and distance of RX~J0852.0-4622 are still debated. \cite{iyudin} estimated 
an age of $\sim680$~yr and the distance of $\sim200$~pc based on the flux of 
$^{44}$Ti lines detected 
with COMPTEL onboard the {\it Compton Gamma-Ray Observatory}. \cite{tsunemi} estimated a 
similar age of between 630 and 970 yr based on observations of Ca X-ray lines with {\it ASCA}. These 
estimates imply that RX~J0852.0-4622 belongs to the supernovae that are closest to Earth. However, \cite{slane} argued that the column density for the X-ray spectrum of the SNR is higher than 
that for the Vela SNR and that the distance to RX~J0852.0-4622 should be much larger, 1–2 kpc. 
On the other hand, \cite{moriguchi} observed the molecular distribution using $^{12}$CO (J=1–0) 
emission measured with the millimeter and submillimeter telescope NANTEN. They estimated the upper 
limit on the distance to be $\sim1$~kpc. Recently, a new estimate of the age and distance has been reported 
from an expansion measurement of the NW rim of the SNR \citep{katsuda}; 1700–4300 yr 
and 0.75 kpc for the age and distance. Assuming a high shock speed of 
3000 km$\cdot$s$^{-1}$, these values seems probable for the SNR. 

\section{Data and analysis}
\subsection{Data reduction}
We used {\it XMM}-{\it Newton} archival data of 14 pointing observations with slightly different aiming 
points. We analyzed data from the European Photon Imaging Camera (EPIC), which consists of 
two MOS and one PN CCD arrays. All observations were performed in the PrimeFullWindow 
mode, either with the medium or the thin filter. The difference between the medium and the thin filter is excluded from the following analysis since we only used a high-energy band of 2--10~keV for spectral 
fitting. Reduction and analysis of the data were performed following the standard procedures using 
the SAS 11.0.0 software package. Since the fluxes measured by the EPIC PN instrument show 
systematically lower values (by 10--15\%) in comparison with the fluxes measured by the EPIC 
MOS, we used only the MOS data. The total exposure time amounts to 436 ks for EPIC-MOS1. The 
summary of the archive data is shown in Table 1.

\subsection{Spectral analysis}
To investigate the spatial properties of the nonthermal emission in the NW rim, we first combined 
different pointing images from EPIC-MOS1 and MOS2 with the {\it emosaic} tool. Although an expansion of the shock front over a time span of 6.5 yr was reported by \cite{katsuda}, we did not 
consider this because the possible displacement ($\sim0.1^{\prime}$) due to the remnant expansion is quite 
small. Figure 1 shows the combined image in an energy range of 0.5-10~keV without exposure
correction. Sharp edges that represent outer shock fronts can be clearly seen in the image. There is also a fainter shock structure located inside the outer boundary (see region ID 6 in Fig. 1). The X-ray emission arises from the interior past of the remnant, even 
$\sim10^{\prime}$ 
from the rim.

The source photons were accumulated from separate rectangular regions with an angular width 
of $0^{\prime}.67$ in the region 1-4 and $1^{\prime}.34$ in the region 5-9 as shown in Fig. 1 (right). We co-added 
source photons of both detectors (MOS1 and 2) to increase the statistics. Since thermal emission 
from the Vela SNR is dominant below 2 keV, we only used 2–10 keV for spectral fitting \citep{hiraga}. All spectra were well reproduced by an absorbed power-law model. Figure 2 shows the 
background-subtracted spectrum of each rectangular region. The absorption column density was 
fixed at $N_{\rm H}=0.58\times10^{22}$ cm$^{-2}$, which was determined from region 1. Figure 3 (top and middle panels) shows the fitting results of the photon index and the flux integrated from 2 keV to 10 keV. 
The photon index shows gradual softening from 2.56 on the rim to 2.96 in the interior region. 
Though radial profiles of the surface brightness have been studied in various shell-type SNRs, a 
gradual softening of the photon index like this has not been seen before. It provides a new clue as to how 
high energy electrons evolve in a shock downstream. The results are summarized in Table 2 (left column). 

\begin{figure*}[htdp]
      \includegraphics[width=18cm]{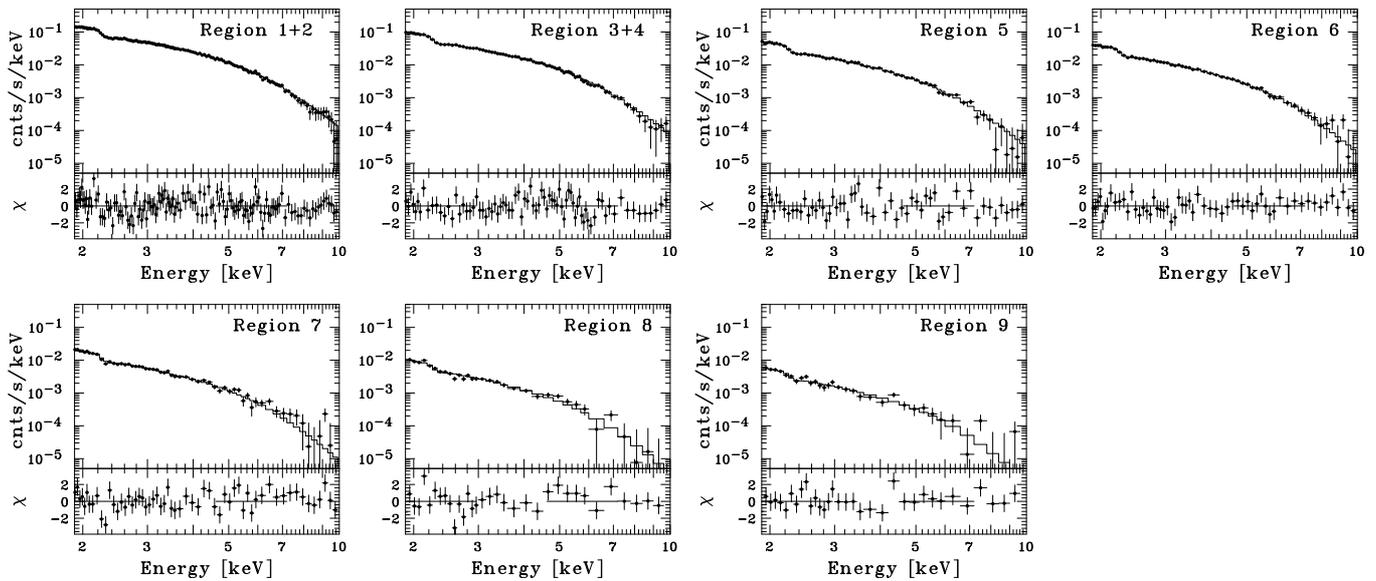}
  \caption{Spectrum of each rectangular region with the best-fit absorbed power-law model. Parameters are 
given in the left column of Table 2. 
}
   \label{fig2} 
\end{figure*}

The nonthermal X-ray spectrum in RX~J0852.0-4622 can be interpreted as synchrotron emission produced by high-energy electrons at a cutoff region of their energy spectrum. The observed 
systematic change of the photon index in the radial direction (Table 2, left column and Fig. 3, 
top panel) is most likely caused by the change of the cutoff position in the synchrotron spectrum. 
Therefore, to investigate this idea, we fit the same data by another model. We employed the following function to fit the X-ray spectrum\footnote[1]{The high-energy cutoff of shock-accelerated electrons is due to the balance between acceleration and synchrotron radiation losses at the shock front if the magnetic field is strong. \cite{zira} have derived the energy spectrum of synchrotron radiation under such circumstances. To allow for other scenarios 
for the formation of the spectral cutoff, we employed Eq. (1).}:
\begin{equation}
F(\epsilon)\propto\epsilon^{-\Gamma}\cdot\exp{[-\bigl( \frac{\epsilon}{\epsilon_0}  \bigr)^{1/2}]},
\end{equation}
where $\Gamma$ is the photon index and $\epsilon_0$ is the cutoff energy. In this approach, our aim is to determine $\epsilon_0$ from the observed variation of $\Gamma$. Therefore the cutoff energy $\epsilon_0$ was treated as a free 
parameter, while the index $\Gamma$ was fixed. This functional form can be regarded roughly as the 
synchrotron radiation produced by electrons with an energy distribution of
\begin{equation}
N(E)\propto E^{-p}\cdot\exp{[-\bigl(\frac{E}{E_{\rm max}}\bigr)]},
\end{equation}
where $E_{\rm max}$ is the maximum energy of electrons. Applying a $\delta$-functional approximation to the synchrotron spectrum of a single-energy electron, we obtain Eq. (1) from Eq. (2) with a relation 
of $\Gamma = ( p + 1)/2$. Moreover, the cutoff energy of the synchrotron spectrum $\epsilon_0$ is related to $E_{\rm max}$ as \citep{aha00, uchiyama03}
\begin{equation}
\epsilon_0\simeq5.3\cdot\bigl(\frac{B}{10~\mu{\rm G}}\bigr)\cdot \bigl(\frac{E_{\rm max}}{100~{\rm TeV}}\bigr)^{2}~{\rm keV}.
\end{equation}

In some young SNRs, a shock-acceleration spectrum with $p 
\simeq 2.2$ has been inferred from 
recent multiwavelength data. Hence, one should expect $\Gamma \simeq1.6$, provided that the power-law part 
of the electron spectrum does not suffer from deformation due to energy-dependent losses. On 
the other hand, if the cooling time of the electrons responsible for the synchrotron X-ray emission 
is shorter than the source age, the volume-integrated spectrum is expected to be characterized by 
a steeper power-law part of $\Gamma\sim2$. We adopted two cases of $\Gamma = 1.6$ and 2.0. The best-fit cutoff energies are presented in Table 2 (middle and right columns) and Figure 3 (bottom panel). Quoted 
errors are at the 1$\sigma$ confidence level. We obtained good fits ($\chi^{2}_{\mu}$, $\mu$) for each region.

\begin{table*}[htdp]
\caption{Summary of the archive data.}
\begin{center}
\begin{tabular}{ccccc}
\hline
\hline 
Obs. ID & Instrument Mode & Filter & Obs. Date & MOS1 GTI (ks) \\ 
 \hline
112870301 &PrimeFullWindow& Medium &2001 Apr. 25 &30.4 \\
137550901 &PrimeFullWindow& Medium &2001 Dec. 10 &34.9 \\
153750701 &PrimeFullWindow& Medium &2002 May 21 &22.3 \\
156960101 &PrimeFullWindow& Medium &2002 Nov. 06 &13.6 \\
159760101 &PrimeFullWindow& Medium &2003 Jun. 22 &17.5 \\
159760201 &PrimeFullWindow& Medium &2004 Nov. 01 &22.4 \\
159760301 &PrimeFullWindow &Thin &2005 Nov. 01 &33.4 \\
162360101 &PrimeFullWindow &Medium &2003 Nov. 10 &23.9 \\
162360601 &PrimeFullWindow &Medium &2003 Nov. 10 &2.1 \\
162363101 &PrimeFullWindow &Medium &2003 Dec. 24 &18.7 \\
412990101 &PrimeFullWindow &Thin &2006 Nov. 09 &58.9 \\
412990201 &PrimeFullWindow &Thin &2007 Oct. 24 &61.2 \\
412990501 &PrimeFullWindow &Thin &2008 Nov. 12 &55.9 \\
412990601 &PrimeFullWindow &Thin &2009 Oct. 25 &40.7 \\
\hline
\end{tabular}
\end{center}
\label{tab1}
\end{table*}%

{\setlength{\tabcolsep}{0.2em}
\begin{table*}[htdp]
\caption{Summary of the spectral fittng.}
\small
\begin{center}
\begin{tabular}{ccccc|ccc|ccc}
\hline
\hline 
\multicolumn{5}{c}{power-law} &  \multicolumn{3}{|c}{exp. power-law ($\Gamma=2.0$)} & \multicolumn{3}{|c}{exp. power-law ($\Gamma=1.6$)} \\
 \hline
 Reg. ID & $N_{\rm H}$ [$\cdot10^{22}$/cm$^{2}$]& $\Gamma$ & Flux (2-10~keV) &$\chi^{2}_{\nu}(\nu)$ & $N_{\rm H}$ & $\epsilon_0$ [keV]   &$\chi^{2}_{\nu}(\nu)$ & $N_{\rm H}$ & $\epsilon_0$ [keV]   &$\chi^{2}_{\nu}(\nu)$ \\
\hline 
1 &0.58 $\pm$0.09 & 2.56 $\pm$ 0.02& 2.74 $\pm$ 0.01 & 1.08 (176) &0.42 $\pm$ 0.08 &3.59 $\pm$ 0.27 &1.03 (176) &0.14 $\pm$ 0.03 &1.36$\pm$ 0.06 &1.06 (176) \\
2 &-- &2.63 $\pm$ 0.02 &3.28 $\pm$ 0.01 &1.12 (197)& -- &2.84 $\pm$ 0.18 &1.13 (197) &--& 1.14 $\pm$ 0.03 &1.09 (197) \\
3 &-- &2.68 $\pm$ 0.02 &2.66 $\pm$ 0.01 &1.03 (167) &-- &2.43 $\pm$ 0.17 &1.06 (167) &-- &1.01 $\pm$ 0.03 &1.02 (167) \\
4 &-- &2.74 $\pm$ 0.03 &2.10 $\pm$ 0.01 &0.93 (119) &--& 2.10 $\pm$ 0.20 &0.98 (119)& --& 0.92 $\pm$ 0.06 &0.92 (119) \\
5 &-- &2.84 $\pm$ 0.02 &1.18 $\pm$ 0.01 &1.19 (59) &-- &1.80 $\pm$ 0.18 &1.21 (59) &--& 0.86 $\pm$ 0.06 &1.12 (59) \\
6 &-- &2.91 $\pm$ 0.03 &0.80 $\pm$ 0.01 &0.89 (59) &-- &1.37 $\pm$ 0.14 &0.82 (59) &--& 0.72 $\pm$ 0.06 &0.80 (59) \\
7 &-- &2.99 $\pm$ 0.05 &0.47 $\pm$ 0.01 &1.14 (58) &-- &1.04 $\pm$ 0.19 &1.20 (58) &--& 0.66 $\pm$ 0.07 &1.17 (58) \\
8 &-- &2.95 $\pm$ 0.09 &0.29 $\pm$ 0.01 &1.2 (30) &-- &1.00 $\pm$ 0.41 &1.24 (30) &--& 0.55 $\pm$ 0.10 &1.22 (30) \\
9 &--& 2.96 $\pm$ 0.22 &0.16 $\pm$ 0.01 &0.90 (32) &--& 0.96$\pm$ 0.63 &0.97 (32)& --& 0.56 $\pm$ 0.12 &0.98 (32) \\
\hline
\end{tabular}
\end{center}
\label{tab2}
\end{table*}%

\begin{figure}[htdp]
      \includegraphics[width=8.5cm]{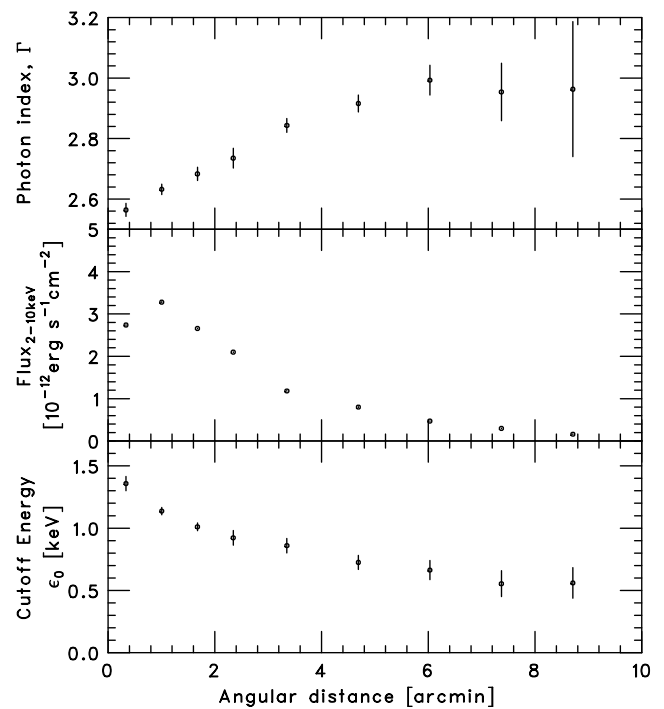}
  \caption{Variation of the photon index (top panel), the flux integrated from 2 keV to 10 keV (middle panel), and the
cutoff energy $\epsilon_0$ obtained from an exponential cutoff power-law model with $\Gamma = 1.6$ (bottom panel). }
   \label{fig3} 
\end{figure}

\begin{figure}[htdp]
      \includegraphics[width=8.5cm]{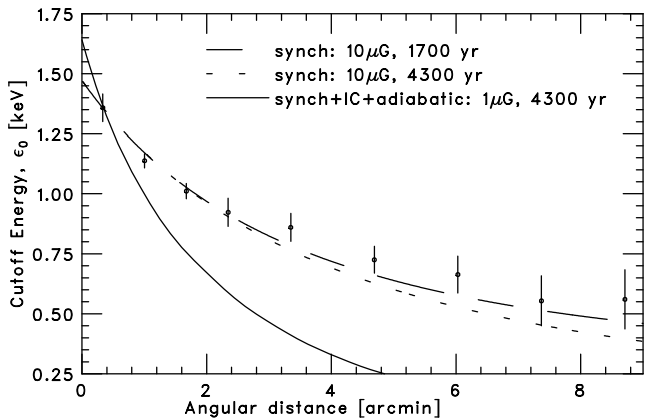}
  \caption{ Spatial dependence of the cutoff energy $\epsilon_0$ obtained from the exponential cutoff power-law model of $\Gamma = 
1.6$ with the time-dependent spectral evolution model.}
   \label{fig4} 
\end{figure}

\begin{figure}[htdp]
      \includegraphics[width=8.5cm]{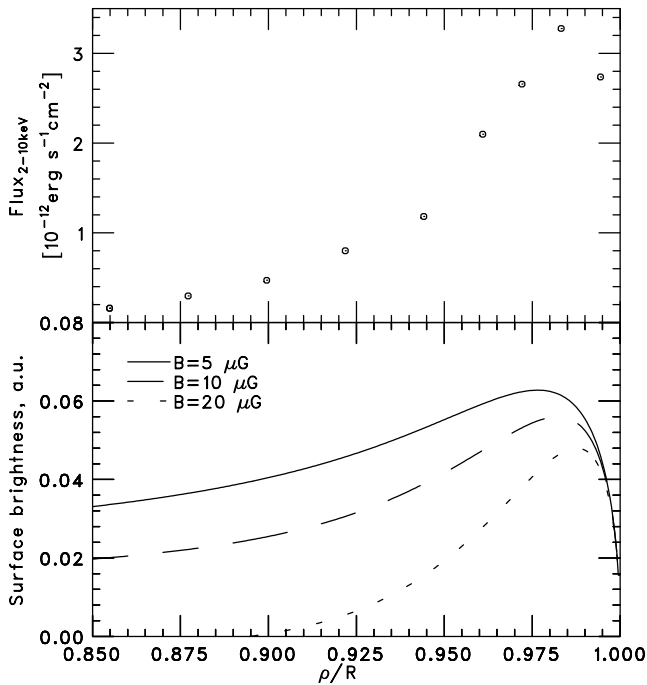}
  \caption{Radial profiles of the observed X-ray flux (top panel) and approximation models (bottom panel), proposed by Petruk et al. (2011). The horizontal axis indicates the normalized distance from the outer shock wave. 
}
   \label{fig5} 
\end{figure}

\section{Discussion}
\subsection{Interpretation of variation in spectral shape}
The fitting results do not show any significant difference except for the absorption column density between simple power-law and exponential cutoff models because of a limited energy band. However, it is still hard to interpret that the electrons are formed in different power-law shapes with region-dependent acceleration efficiencies behind the shock front. This assumption comes from the widely 
accepted particle-acceleration theory, i.e., the diffusive shock acceleration or the first-order Fermi 
acceleration, which specifies the effective acceleration site at the shock front with a power-law electron spectrum with an index of $\sim2$. Therefore the different photon indices can be more naturally 
interpreted as the presence of spectral cutoffs associated with the unavoidable cutoff in the acceleration spectrum of parent electrons. A solid evidence of such cutoff energy has been reported in a 
young SNR RX~J1713.7-3946 (on the order of 1000 yr) with the {\it Suzaku} broadband spectrum of more than two 
decades in the energy band, i.e., $\sim40$~keV \citep{tanaka08}. Although the FoV of the {\it Suzaku} HXD 
is not small enough to allow detailed imaging spectroscopy, \cite{tanaka08}'s study suggests that a spectral 
cutoff is a common feature for the synchrotron spectrum, which shows a typical photon index of 
2-3 below 10~keV of shell-type SNRs. We note that the obtained absorption column density also 
indirectly suggests the presence of the energy cutoff in the spectrum of the Vela Jr. \cite{slane} pointed out that the column density for the power-law is significantly higher than that for Vela 
SNR, e.g., $\sim10^{20}$ cm$^{-2}$ \citep{bocchi, asch95}. If we assume the distance to the Vela Jr. to be 0.75 kpc, the lower column density with 
the exponential cutoff model, i.e., $1.4\times10^{21}$ cm$^{-2}$, seems physically preferable in comparison with 
a distance to the Vela SNR of 
$\sim$0.25~kpc \citep{cha}. We thus interpret the 
different photon indices as energy cutoffs of the parent electron spectra in the following discussion.

\subsection{Synchrotron cooling effect on the cutoff energy}
The variation of the cutoff energy reflects the time evolution of the electron spectrum. Once a particle distribution has been injected at the shock, it evolves downstream, suffering from radiative losses 
and adiabatic expansion losses. We first ignored the adiabatic loss and focused on the synchrotron 
cooling effect on the cutoff energies to estimate an upper limit of the magnetic field strength. We consider the following equation to describe a time evolution of electrons in a certain volume 
element only subject to synchrotron losses: 
\begin{equation}
\frac{\partial N(E)}{\partial t}=\frac{\partial}{\partial E}[b(E)N(E)]+Q(E,t),
\end{equation}
where $t$ denotes time elapsed since the shock, $b(E) = aE^{2}$ is the energy loss rate of the electrons, namely, $b(E)=-\bigl(\frac{dE}{dt}\bigr)$, and $Q(E)$ is a source term that describes the rate of injection of electrons and their injection spectrum into the source region. If we assume the injection of electrons with a power-law energy spectrum at $t = 0$ without subsequent injection of electrons, we can write the electron 
spectrum as $Q(E) = \kappa E^{-p} \delta(t)$, where  $\delta(t)$ is the Dirac delta function. It is straightforward to show 
that the solution of Eq. (4) is given as \citep{longair} 
 \begin{equation}
N(E,t)=Q_0E^{-p}  \cdot (1-atE)^{p-2} \cdot  \Theta\bigl(  \frac{1}{at}-E \bigr),
 \end{equation}
 where $ \Theta$ is a step function. The factor $(1-atE)^{p-2}$ represents a cutoff shape at $E_{\rm cut}=(at)^{-1}$ due to synchrotron losses. We may interpret $E_{\rm max}$ of Eq. (2) as $E_{\rm cut}$. However, since the injection spectrum 
also has a cutoff in reality, this interpretation would not be always appropriate.

If we assume that the downstream evolution of $E_{\rm max}$ is simply determined by synchrotron 
losses, we can obtain the time evolution of the electron energy by integrating the energy loss function, i.e., $\int-\frac{dE}{E^{2}}=\int a \cdot dt$, and then, 
\begin{equation}
E_{\rm max}=(at+E_0^{-1})^{-1}, 
\end{equation}
where $E_0\equiv E_{\rm max} (t=0)$. The radial evolution of $\epsilon_0$ can be calculated from $E_{\rm max}(t)$ with a transformation of $t$ into $x$, which is the distance from the shock to the volume element. We assume that the 
shock velocity has a time dependence of $V \propto T^{-3/5}$ 
(the Sedov-Taylor evolution), where $T$ 
denotes the age of the remnant. The coefficient of the synchrotron energy loss rate $a$ is given as
\begin{eqnarray}
b(E)&=&aE^{2}\nonumber\\
&=&6.6\times 10^{4}\gamma^{2}B^{2}~[{\rm eV/s}] \nonumber\\
&=&2.52\times\bigl( \frac{B}{10~\mu{\rm G}}\bigr)^{2}\cdot\bigl(\frac{E}{100~{\rm TeV}}\bigr)^{2}~[{\rm keV/s}],
\end{eqnarray}
where $\gamma$ is the Lorentz factor of the electron, and $B$ is magnetic field strength. Defining $k = a/B^2$ , Eq. (6) can be written as 
\begin{equation}
E_{\rm max}(x)\propto[kB^{2}\cdot t + E_0^{-1}]^{-1},
\end{equation}
where $B$ is the magnetic field strength (assumed to be constant in $t$, i.e., $B$ is uniform downstream, 
as well as $T$ , i.e., $B$ is uniform upstream of the forward shock). From Eq. (3) we can obtain 
the cutoff photon energy as
\begin{equation}
\epsilon_0(x)\propto B\cdot[kB^2\cdot t +E_0^{-1}]^{-2}.
\end{equation}

We adopt a current shock velocity of $V = 3000$~km$\cdot$s$^{-1}$ and distance of $D = 0.75$~kpc \citep{katsuda}. The coeffcient $E_0$ is chosen so that the spatial curves pass the initial data point. Then, 
the magnetic field strength controls the spatial variation of the cutoff energy. The time-dependence 
of the shock velocity, i.e., the Sedov-Taylor evolution, is included in the model. We used $V_{\rm rel} = (1/4)V$ 
as a relative speed between the shock front and post-shock gas to transform $t$ into $x$. Figure 4 shows 
the spatial dependence of the cutoff energy with the time-dependent spectral evolution model. The 
magnetic field strength of $B \sim10 \mu$G can reproduce the decrease of $\epsilon_0$ as a function of $x$ under 
the assumption that the field strength does not change both in time and space. We note that a 
projection effect, which is neglected in our analysis, dilutes the spatial variation of $\epsilon_0$, and 
consequently could increase the required magnetic field strength somewhat.

\subsection{Adiabatic and inverse Compton scattering effects on the cutoff energy}
The electrons would significantly suffer from the adiabatic losses. In this case, the required magnetic field strength becomes lower than the above estimated value, and then the inverse Compton 
scattering effect also needs to be considered. The adiabatic loss rate for the electrons 
due to expansion of the volume is given as (Longair, 1994)

\begin{equation}
\bigl(\frac{dE}{dt}\bigr)_{\rm adiabatic}=-\frac{1}{R}\cdot\bigl( \frac{da}{dt}\bigr)\cdot E,
\end{equation}
where $a$ is the fluid coordinate along the radius of the remnant $R$. Here we again assume that an 
electron population has been confined within a region of certain volume and injected at the shock in 
a fluid form. The energetic electron fluid does work adiabatically as it expands, and consequently 
loses its internal energy.

As for the inverse Compton scattering, the seed photon fields include the cosmic microwave 
background (CMB) and Galactic infrared and optical fields in the interstellar medium. In general, 
the photon fields produced locally by the SNR itself would not contribute to the target soft photons 
and Compton scattering of CMB photons should dominate the others. In the Thomson regime, the 
energy-loss rate via inverse Compton scattering of the CMB field is written as

\begin{equation}
\bigl(\frac{dE}{dt}\bigr)_{\rm IC}=-\frac{4}{3}c\sigma_{\rm T}\beta^{2}\gamma^{2}U_{\rm CMB},
\end{equation}
where $\sigma_{\rm T}$ is the Thomson cross-section, $\beta$ is the velocity of an energetic electron in units of the speed of light $c$, and $U_{\rm CMB} \simeq 0.26$~eV cm$^{-3}$ 
is the energy density of CMB photons. We 
initially assumed an input electron spectrum as an exponential power-law shape with $p = 2.2$ and 
numerically calculated the electron spectrum after losing energy due to the synchrotron, adiabatic, 
and inverse Compton scattering effects. The cutoff energies were obtained by fitting the synchrotron 
spectra produced by the continuously energy-modulated electron spectra. We assumed a uniform 
interstellar magnetic field strength of 
$\sim1~\mu{\rm G}$. As is shown in Fig. 4, the adiabatic effect dominates 
over the synchrotron cooling and the cutoff variation can be explained without radiation losses. 
In the adiabatic loss-dominant case, a site of the magnetic field amplification might be focused 
on the bright filament region and the field strength of the inner regions is similar to a normal 
interstellar magnetic field. The magnetic field strength of the filament is estimated as 
$\sim6~\mu{\rm G}$ from 
the X-ray/TeV observations and the leptonic model \citep{aha07}. This implies 
that the magnetic field amplification near the shock occurs only moderately in the SNR. However, Eq. 
(10) might be a too simplified assumption because the electron population might not be suitable as an ideal expanding fluid for over 1 kyr, e.g., at a post shock flow speed on the order of $3000/4$ km$\cdot$s$^{-1}$, 
electrons fall behind the shock $1^{\prime}$ or $6.74\times10^{12}$ 
km in 
$\sim300$ yr. Thus we conclude that the magnetic 
field strength near the filament is estimated to be between $6$--$10~\mu$G. 

\subsection{Application of MHD-based model to the radial profile}
To confirm the consistency of the above estimated magnetic field strength with an MHD- 
based calculation, we applied the analytical model proposed by \cite{pet11} to the obtained 
radial profile. This model accounts for the projection and other MHD effects, such as a variation of 
the magnetic field strength in the post-shock region, and calculates synchrotron images of SNRs of 
the Sedov-Taylor phase in a uniform interstellar medium. Figure 5 shows the comparison between the 
observed radial profile of the X-ray flux and the model curves. We assumed the cutoff energy near 
the shock front to be $\epsilon_0 = 4$~keV and the shock obliquity angle $\Theta_0= 0^{\circ}$, which is defined as the angle 
between the magnetic field and the normal to the shock. For other parameters, we used typical values 
as follows: $\kappa_{\rm ad} = 1$, $\kappa_{\rm r} = 8.25$, $\sigma = 4$, $\gamma = 5/3$, $b = 0$, $q = 0$, $t = 1000$ yr, $\epsilon_{\rm keV} = 2$~keV (see Petruk 
et al. 2009, 2011 for the definitions of the model parameters).

We calculated radial profiles for three sets of ( $B$, $E_{\rm max}$ ): (i) $B = 5~\mu$G and $E_{\rm max}= 122.8$~TeV, 
(ii) $B = 10~\mu$G and $E_{\rm max} = 87$~TeV, and (iii) $B = 20~\mu$G and $E_{\rm max} = 61.4$~TeV. Each set gives $\epsilon_0 = 4$~keV using Eq. (3). The angular radius of the SNR was adopted to be $R = 1^{\circ}$. The radial profile of the 
X-ray flux is broadly consistent with the models with $B = 5$--$20~\mu$G, which roughly agrees with the 
magnetic field used above to reproduce the spatial variation of $\epsilon_0$.

If the $\gamma$-ray emission detected by {\it Fermi} and H.E.S.S. is predominantly caused by IC scattering, 
the magnetic field in the $\gamma$-ray production region should be 
$\sim10 \mu$G because of the measured ratio of 
the TeV $\gamma$-ray and X-ray energy fluxes, $\omega_{\gamma}$ (1–10 TeV)/$\omega_{\rm X}$ (2–10 KeV) $ \sim2$ \citep{aha05}.
The magnetic field required to explain the radial profile would favor the leptonic model. However, 
this is not conclusive because the observational constraints are not sufficient to explore 
other possibilities that may account for the variation of $\epsilon_0$. More detailed multiwavelength observations 
are necessary to place more definitive constraints on the magnetic field and to elucidate the 
$\gamma$-ray emission mechanism.

\end{document}